\RequirePackage{fix-cm}
\documentclass[smallextended]{svjour3}       
\smartqed  
\usepackage{graphicx}
\usepackage{amsmath}
\usepackage{amssymb}
\usepackage{mathrsfs}
\usepackage{bm}
\usepackage{natbib}
\usepackage{color}
\usepackage{tikz}
\usepackage{eucal}[mathcal]
\usepackage{siunitx}
\usepackage[colorinlistoftodos]{todonotes}

\usepackage[pdftex,bookmarks]{hyperref}

\usepackage{dcolumn}
\newcolumntype{d}[1]{D{.}{.}{#1}}

\newcommand{\eotvos}{E$\ddot{\rm o}$tv$\ddot{\rm o}$s}
\newcommand{\eotwash}{E$\ddot{\rm o}$t-Wash}

\allowdisplaybreaks
\begin{document}

\title{On the first test of the Weak Equivalence Principle in low Earth orbit}

\author{Anna M. Nobili \and
 Alberto Anselmi
}


\institute{A.\,M. Nobili\at
  University of Pisa, Department of Physics ``E. Fermi'',
  Largo B. Pontecorvo 5, 56127 Pisa, Italy\\
  Tel.: +39-050-2213254\\
  \email{anna.nobili@unipi.it}
  \and
 A. Anselmi\at
Thales Alenia Space Italia, Strada Antica di Collegno 253, 10146 Torino, Italy (ret.)
}

\date{Received: date / Accepted: date}

\maketitle

\begin{abstract}
 The Weak Equivalence Principle is the founding pillar of General Relativity and as such it should be verified as precisely as possible. The \textit{Microscope} experiment tested it in low-Earth orbit, finding that Pt and Ti test masses fall toward Earth with the same acceleration to about 1 part in $10^{15}$, an improvement of about two orders of magnitude over ground tests. Space missions, even if small, are expensive and hard to replicate; yet, the essence of physics is repeatability. This work is an assessment of the \textit{Microscope} results based on the laws of physics and knowledge from previous experiments, focusing on the limiting thermal noise and the treatment of acceleration outliers. Thermal noise reveals anomalies that we explain by stray sub-$\mu\rm V$ potentials caused by patch charges, giving rise to an unstable zero. The measurements were affected  by numerous acceleration spikes occurring at the synodic frequencies relative to the Earth (the signal frequency) and the Sun, which we interpret as evidence of a thermal origin. In \textit{Microscope} authors’ analysis, the spikes were removed and the resulting gaps replaced with artificial data (up to $35\%$, $40\%$ of the sessions data), which retain memory of the gaps and  may simulate or cancel an effect (signal or systematic). An alternative approach based exclusively on real measured data would avoid any ambiguity. The lessons of \textit{Microscope} are crucial to any future improved mission.
  
\end{abstract}

\section{Introduction}\label{Sec:Introduction}

The Weak Equivalence Principle (WEP), or rather not a ``principle'' but  the ``hypothesis'' of a complete physical equivalence of the gravitational field and the corresponding acceleration of the reference frame --as Einstein expressed it in 1907-- is the founding pillar of General Relativity. In Newton's formulation it was the equivalence of inertial to gravitational mass and the Universality of Free Fall (UFF) (\cite{AJP-2013}). As such, it should be tested as precisely and accurately as possible, something  physicists have been doing since  Galileo's time.

To date, the best limit in laboratory controlled experiments has been established by the rotating torsion balances of the \textit{\eotwash\ }group, achieving a fractional accuracy of about $10^{-13}$ in the field of the Earth for $\rm Be$, $\rm Ti$ and $\rm Al$, $\rm Pt$ test masses (\cite{Adel2008,AdelCQGfocusissue2012}). Recently, the \textit{Microscope} mission has tested the WEP  for the first time  in a low Earth orbit. By exploiting  a driving signal almost 500 times stronger than on the ground (Sec.\,\ref{Sec:AccelerometersSignal}), \textit{Microscope} has improved the limit by about two orders of magnitude, down to about $10^{-15}$, for $\rm Pt$ and $\rm Ti$ test masses (\cite{MS-PRL2022}).

Even a small space mission like \textit{Microscope }($300\,\rm kg$) is a ``big science'' project hard to replicate. Yet, the essence of science is repeatability.
The mission was carried out by a single team as a single experiment. The measurement data are publicly available, but the practice of having two independent teams performing the data analysis in parallel (e.g.\ as in the \textit{Hipparcos} mission of ESA) has not been pursued. On the other hand, a new data analysis performed \textit{post facto,} without the in depth knowledge (of the experimental setup, spacecraft, software and operations) that comes from first hand participation in a satellite mission, would be highly controversial and ultimately inconclusive --if it ended up with anything inconsistent with the original results-- or irrelevant --if it followed too closely the steps of the first analysis.

Does this mean that the result of \textit{Microscope} shall be fully assessed only by another space mission of comparable or higher sensitivity? The answer is no,  because it is possible to break down the experiment  into its key features, each of which  can be investigated on the basis of basic physics arguments and well established knowledge, both theoretical and experimental. 

The basic features of the \textit{Microscope} experiment are addressed in Secs.\,\ref{Sec:AccelerometersSignal} and\,\ref{Sec:ForcesControl}. 
Sec.\,\ref{Sec:AccelerometersSignal} is concerned with the test masses, the accelerometers and the acceleration measurements, a potential violation signal from Earth, its signature and how it would be measured by a Sensor Unit (SU). In particular we show the $500$ gain factor in space and stress the difference  between an inherently differential instrument like the torsion balance (no violation, no signal) and the SU of \textit{Microscope} in which the acceleration difference is reconstructed from the individual measurements of two independent accelerometers. The relevance and consequences of  such feature are  addressed in Secs.\,\ref{Sec:ForcesControl} and \ref{Sec:LimitingNoiseAndFluctuatingZero}.

Sec.\,\ref{Sec:ForcesControl} deals with the forces acting on the test masses, starting from drag and drag-free control and focusing on the electrostatic control of the test mass in the presence of Earth tidal forces, mechanical stiffness, biases, patch charges and fluctuating patch potentials. 

Establishing the nature and source of the noise limiting the measurement is crucial for any experiment and a prerequisite for future improvements. 
    In Sec.\,\ref{Sec:LimitingNoiseAndFluctuatingZero} we analyze the limiting random noise. Since the preliminary results (\cite{MS-PRL2017}) the residual acceleration difference was reported to be limited by  thermal noise  with  $1/\sqrt{\nu}$ frequency dependence  attributed to  internal damping in the gold wires. However, our analysis of the final results for all measurement sessions,  which appeared at the same time  in \cite{MS-PRL2022} and in a series of papers  by the \textit{Microscope} scientists on volume 39 of Classical and Quantum Gravity, 
reveals anomalies which are inexplicable within the current best knowledge of this noise. The GOCE mission (see \cite{GOCE-2010} for a general description) has shown that fluctuating patch potentials with $1/\sqrt\nu$ trend  at low frequency were one of the main noise sources. We show that such a  potential at sub-$\mu\rm V$ level  at the frequency of the signal would be sufficient to explain the observed anomalies. This is because the fluctuating voltage multiplies a considerably large voltage needed to keep the test mass motionless at an arbitrary chosen zero of the sensor. As a result, a large force bias arises, hence, by way of the fluctuating patch potentials, a fluctuating zero. An unstable zero is a known issue  in WEP tests, related to the problem of initial conditions (or release) errors,  affecting all tests  not  designed as intrinsically null experiments. 

The presence of patch charges and consequent fluctuating potentials at the frequency of the signal should be revisited in order to firmly establish its contribution to the limiting noise, starting from the knowledge acquired with GOCE. 

Systematic effects that were a matter of concern (\cite{MicroscopeLimitations}) turned out to be dominated by temperature variations. After calibration of the thermal response and \textit{a posteriori }correction,  residual systematic errors were much smaller than random errors or close to them (\cite{TouboulCQG2022-9, RodriguesSystematicsCQG2022})  
This was achieved thanks to an effective multilayer thermal insulation blanket (MLI). However, MLI blankets are the likely source 
of  large acceleration spikes  (``glitches'') occurring simultaneously in the four test masses (Sec.\,\ref{Sec:OutliersGapsArtificialData}). GOCE  was exempt thanks to an \textit{ad hoc}, stiffer than usual, solution adopted for the insulation. 

\textit{Microscope's} glitches appear to be induced by thermal stress from the Earth and the Sun on the MLI. They occur at the synodic frequency relative to Earth (also the frequency of a violation signal) and the Sun, their  harmonics and their difference. Outliers were removed and the ensuing gaps  (up to $35\%$, $40\%$ of the sessions data) were filled with artificial reconstructed data. However, gaps and artificial data retain memory of the outliers they have replaced and may therefore mimic a violation signal or cancel an effect (signal or systematic).

An alternative approach relying solely on real measured data has been successfully used in  the \textit{E\"ot-Wash} ground  tests, with only $7\%$ of data removed (\cite{Adel2008}). We argue that the same procedure is applicable to the \textit{Microscope} test in orbit. 

Conclusions are drawn in Sec.\,\ref{Sec:Conclusions}, along with the main takeaway  lessons for  future improvements.


\section{The accelerometers of Microscope and the signature  of a potential WEP violation signal}\label{Sec:AccelerometersSignal}

Inside an Earth orbiting spacecraft its center of mass (CoM) is the only zero force point  (perfect weightlessness) since in this point the Earth's gravitational attraction is perfectly balanced by the inertial centrifugal force arising in the non inertial reference frame of the spacecraft. Any Test Mass (TM) located at a non zero position vector relative to the CoM of the spacecraft is subjected to a tidal acceleration because the gravitational field is not uniform, resulting in gravity gradient forces referred to as  tidal forces  (since ancient times sea tides were attributed to the Moon and the Sun because of their typical frequencies).  

If a TM inside an Earth orbiting spacecraft  violates the equivalence between its inertial and gravitational mass ($m_1^i$ and $m_1^g$ respectively) it is $m_1^g=m_1^i(1+\eta_{_1})$ where $\eta_1\neq0$ is a  dimensionless parameter quantifying the violation. In this case the Earth acts on the TM not only through the tidal acceleration, but also with an additional acceleration $a_{_1}=g\eta_{_1}$ where $g=GM_\oplus/r^2$ is the monopole gravitational attraction from the Earth (with $M_\oplus$ the mass of the Earth, $r$ the orbital distance of the spacecraft, $G$  the universal constant of gravity). If another TM of different composition violates the equivalence of inertial to gravitational mass at level $\eta_{_2}$, and the two masses are perfectly concentric (hence they are affected by the same tidal acceleration), there will be a differential acceleration between the two given by:
\begin{equation}\label{Eq:aEP}
a=a_{_2}-a_{_1}=g\eta\ \ \ \ \  , \ \ \ \   \eta=\eta_{_2}-\eta_{_1} \ \ .
\end{equation}
Thus, the two TMs fall towards the Earth with different accelerations, violating UFF;  $g$ is the ``driving signal'' of  the violation in the field  of the Earth and  the  dimensionless parameter $\eta=a/g$ is the fractional differential acceleration  between the two TMs that quantifies the level of violation. It is named after the Hungarian physicist   \eotvos\  who first designed an intrinsically differential sensor --the torsion balance-- to test the WEP. 

The concepts of equivalence between inertial and gravitational mass, UFF and WEP can all be used to mean the same thing (\cite{AJP-2013}).  
The theoretical framework is that of a 2-body problem (TM in the field of the source body)  with WEP violation,  and the violation signal lies in the orbit plane of the spacecraft (\cite{GRG-2008}). Since the sensor is rigid with the spacecraft it is natural to work in  the non inertial reference frame of the spacecraft centered on its CoM relative to which --in case of space fixed attitude-- the Earth orbits at angular velocity of modulus $\omega_{orb}=\sqrt{GM_\oplus/r^3}$. Since the signal has two degrees of freedom it is desirable that the sensor too be sensitive in 2D (the orbit plane). The advantages over 1D sensors  is to double the measurement data (and thus double the integration time  $t_{int}$ and reduce the random noise by $\sqrt{2}$,  since it is proportional to $1/\sqrt{t_{int}}$), and to avoid the risk of noise leakage from the other degree of freedom  which  in 1D sensors is less sensitive by design.

From Eq.(\ref{Eq:aEP}) it is apparent that, for a given sensitivity of the experiment to differential accelerations between the TMs, the larger the driving signal, the smaller the resulting value of $\eta$, the better the WEP test. 

At the orbiting altitude of \textit{Microscope} ($\simeq700\,\rm km$) it is $g=7.9\,\rm ms^{-2}$. For test masses suspended on the ground, as in a pendulum or in torsion balances, the driving signal is given by the inertial centrifugal acceleration $\omega_\oplus^2R_\oplus\cos{\vartheta}\sin	{\vartheta}$ due to the diurnal rotation of the Earth $\omega_\oplus=2\pi/86164\,\rm rad/s$ which balances the component of the gravitational attraction of the Earth in the horizontal plane tangent to its surface ($R_\oplus\simeq6.38\times10^6\,\rm m$  the Earth radius)  at a given latitude $\vartheta$. Its maximum value (at $45^\circ$ latitude) is $\simeq0.0169\,\rm ms^{-2}$. 

Thus, the \textit{Microscope} test of WEP has almost a 500 factor gain over the  \textit{\eotwash} torsion balance tests (operating close to  $45^\circ$ latitude)  simply by being performed in space. This is in fact the first and foremost advantage (although not the only one) of testing the WEP in space. However, a stronger driving signal does not automatically mean a  correspondingly better WEP test. Its actual achievement depends on experiment design, sensitivity to differential accelerations,  systematic errors, random noise and integration time. For instance, mass dropping tests on the ground can count on a driving signal about 600 times stronger than torsion balances ($9.8/0.0169\simeq580$), but their performance has not yet even reached that of torsion balances, mostly because of initial condition systematic errors. The best mass dropping test has been performed by  atom interferometry (\cite{KasevichWEP-2020}) finding  no violation  to about $10^{-12}$ for $\rm ^{85}Rb$  and $\rm ^{87}Rb$ atoms. With a composition difference by two neutrons only, expectations of  a possible violation are however rather feeble.

Any non zero offset between the centers of mass  of the TMs results in a differential tidal acceleration directly proportional to the offset. In order to reduce the offset, and also for practical reasons, the TMs of a WEP test in space are designed as two concentric, coaxial, hollow cylinders.

\begin{figure}
\begin{center}
\includegraphics[width=0.4\textwidth]{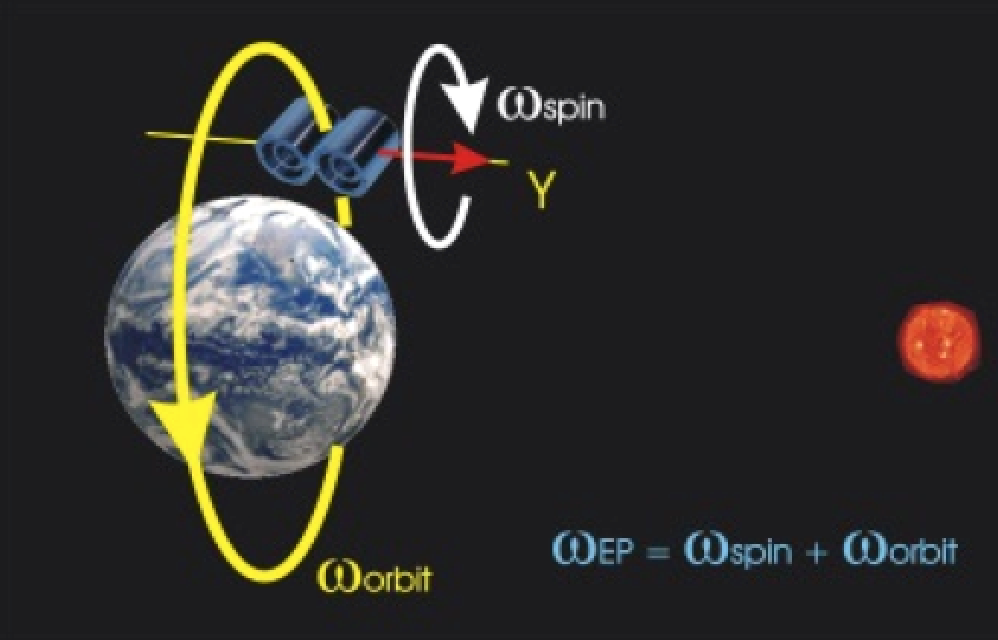}
\hspace{2mm}
\includegraphics[width=0.26\textwidth]{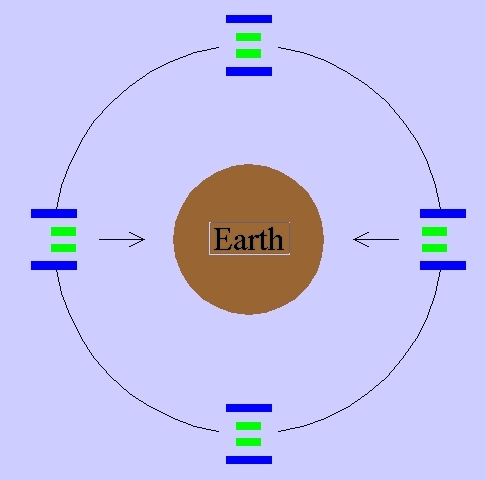}
\caption{The two Sensor Units of \textit{Microscope} in orbit  (left, {figure taken from \textit{Microscope} presentations}) and the WEP violation signal at zero spin (right); figures not to scale. The orbit ($\simeq700\,\rm km$ altitude) is near-circular  ($e\simeq10^{-4}$) and near-polar ($i\simeq98.2^\circ$), making it sun-synchronous with solar panels facing the Sun and the descending node at dawn. In the reference frame of the spacecraft, $Y$ axis points in the direction of the outward orbit normal and 
the $X, Z$  axes lie in the orbit plane. The satellite spins clockwise (opposite to the orbital motion) so that the signal frequency is the sum of the orbit and spin rate. The separation between the centers of mass  of the two Sensor Units is along the $Y$ axis, so that, projected on the $X, Z$ orbit plane, the position of the centers of mass  of the test cylinders can be closer to the  CoM of the satellite, as constrained by construction and mounting errors. At zero spin a WEP violation signal would act along the symmetry axis at  $\omega_{orb}$    (right sketch, section of SUEP sensor on the orbit plane with the different composition cylinders depicted in different colors).}
\label{Fig:SensorUnitsInOrbit}
\end{center}
\end{figure}
\begin{figure}
\begin{center}
\includegraphics[width=0.3\textwidth]{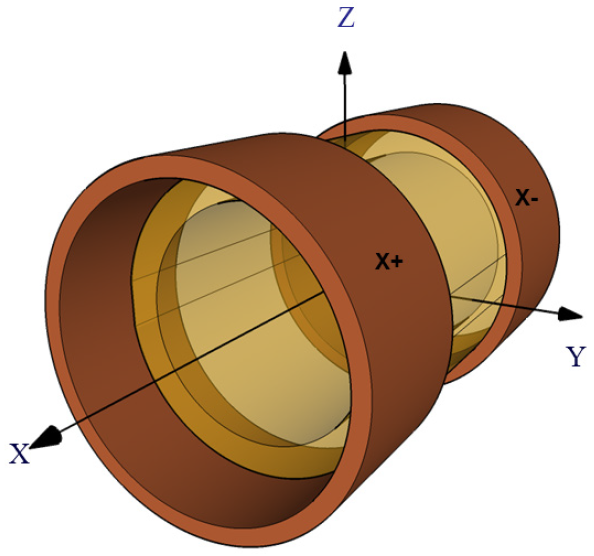}
\caption{\textit{Microscope} test cylinder with $X$ axis electrodes (Fig.5 in \cite{CQG2019}). In each accelerometer two conductive electrode belts are placed at the two ends of the cylinder with a gap of $598\,\mu\rm m$. If the CoM of the cylinder moves away from the central position, the area of the cylinder facing its electrode increases on one side and decreases on the other (``area-variation''), and so does the capacitance, by an amount depending on the displacement, hence on the acceleration that generated it. }
\label{Fig:AreaVariationElectrodes}
\end{center}
\end{figure}

In \textit{Microscope} each test cylinder forms an independent accelerometer; it is suspended electrostatically, its symmetry axis $X$ is the best sensitive axis, it lies in the $X, Z$ orbit plane   and it is devoted to the detection of a potential violation signal. As far as  WEP testing is concerned it is therefore a 1D sensor. However, each test cylinder has 6 degrees of freedom, and must be controlled in all of them. In this work we focus on the $X$ sensitive axis.

Two coaxial accelerometers form what is called a Sensor Unit (SU), and the spacecraft carries two  of them (Fig.\,\ref{Fig:SensorUnitsInOrbit}), one with different composition $\rm Pt$-$\rm Ti$ test cylinders for testing the WEP, named SUEP, and the other with $\rm Pt$-$\rm Pt$ cylinders named SUREF as a reference sensor, since WEP violation --if any-- is expected to occur if the test bodies have different compositions (which may couple differently to the gravitational field of the source body), and not simply different masses.

In each accelerometer two conductive electrode belts are placed around the two ends of the cylinder and measure the capacitance variations  along the  symmetry $X$ axis according to the ``area-variation" scheme  (Fig.\,\ref{Fig:AreaVariationElectrodes}).

A tiny gold wire parallel to the symmetry axis connects the test cylinder to the cage enclosing it  (rigid with the spacecraft) for electric grounding and polarization of the test mass.

Like in all accelerometers by ONERA, the test masses of \textit{Microscope} --although in the shape of hollow cylinders rather than cubes or parallelepipeds-- are not allowed to move. A control voltage is applied to nullify any displacement  so as to keep the TM motionless (except for the residual noise that cannot be controlled) at a chosen capacitive zero $O_C$ (which depends on geometry, construction, mounting and electronics). By calibration, a scale factor is established (in $\rm ms^{-2}/V$) by which the applied electric potential yields the acceleration acting on the test cylinder. Then,  in  \textit{Microscope}   a scale matching  is performed between the two accelerometers of each SU and  their acceleration difference is obtained, which is the physical quantity of interest for testing the WEP. This shows that the SU is not an intrinsically  differential sensor, meaning that it does not measure the differential acceleration of the test masses directly, but provides it as the difference of two separate  acceleration measurements (Secs.\,\ref{Sec:ForcesControl} and\,\ref{Sec:LimitingNoiseAndFluctuatingZero}).

If the spacecraft attitude is fixed relative to inertial space the signal of a WEP violation from the Earth on the sensitive axis of SUEP sensor is an acceleration difference at the orbital frequency and phase as shown in Fig.\,\ref{Fig:SensorUnitsInOrbit}, right sketch, of unknown amplitude and sign.
If the spacecraft spins relative to inertial space around the $Y$ axis perpendicular to the orbit plane  as shown in Fig.\,\ref{Fig:SensorUnitsInOrbit}, a potential violation signal is up-converted  to the higher orbital plus spin frequency (the signs are opposite). Of course, any other effect at the orbital  frequency, such as drag, is up-converted too. 
 
 The advantage  of rotation is to reduce the noise  which is known to decrease at higher frequency (Sec.\,\ref{Sec:LimitingNoiseAndFluctuatingZero}) and also to average thermal effects. However, as shown in both sketches of Fig.\,\ref{Fig:SensorUnitsInOrbit},  by the very design of the \textit{Microscope} sensor, rotation cannot occur around the symmetry axis, which would ensure a higher spin rate and  passive rotation by conservation of angular momentum, in which case attitude manoeuvers are needed only  to keep the solar panels close to the perpendicular to the Sun.

\section{Forces acting on Microscope accelerometers and their control}\label{Sec:ForcesControl}

Residual air along the orbit  and solar radiation impinging on the outer surface of the spacecraft give rise to a force on its CoM we will globally refer to as  ``drag''. As a result,  the reference frame of the spacecraft in which measurements  are performed is an accelerated non inertial frame and any TM inside it is subjected to an inertial acceleration equal and opposite to the non gravitational acceleration of the CoM of the spacecraft.

Ideally, being the same on all TMs (common mode effect) it should not affect the differential acceleration between the two TMs of the sensor unit. In reality, the two accelerometers of the SU do not respond with exactly the same acceleration. Only a limited rejection  of common mode effects is achieved by scale matching in orbit. The residual differential effect of drag has the same frequency as the target signal and almost $90^\circ$ phase difference (at \textit{Microscope's}  low altitude, residual air dominates over solar radiation and the effect of drag is mostly along track, while a violation signal would point towards or away from Earth).
 A  remaining differential component --after drag free control and partial rejection--  with the same frequency and phase as the signal would mimic a violation signal; it is therefore necessary to ensure that the entire residual differential effect be smaller than the target.

The drag free control system receives the non-gravitational acceleration of the spacecraft measured by the accelerometers and synthesizes the commands to the actuators (thrusters) that result in equal forces of opposite sign (with limited precision, in a limited control bandwidth).
In general, one of the two SUs is used to drive the drag free control while the other takes science data. 

In \textit{Microscope} the acceleration  due to drag is about $10^{-8}\,\rm ms^{-2}$. The requirement placed on drag free control is to partially compensate it leaving a residual common mode acceleration of about $10^{-12}\,\rm ms^{-2}$. This would be  then partially rejected by matching in orbit the scale factors of the two accelerometers of each SU such that the residual acceleration difference caused by  drag is reduced to $3$-$4\times10^{-15}\,\rm ms^{-2}$, with a WEP violation target of  $7.9\times10^{-15}\,\rm ms^{-2}$.

The test cylinders of  \textit{Microscope's} accelerometers are not free masses. They are controlled electrostatically, and they are subject to    electrostatic stiffness and to the mechanical stiffness of the gold wire. 

In most ONERA accelerometers (e.g.\,in GOCE) as well as in LISA, the TM is a cube (or a parallelepiped) and displacements are detected via capacitance variations in the classical ``gap-variation'' scheme in which the TM is located halfway between two capacitance plates facing its two opposite faces perpendicular to the sensitive axis. This scheme is used   in \textit{Microscope} for the $Y, Z$ axes, while the $X$  symmetry/sensitive axis is based on area-variation.
As shown in \cite{ONERA1999},  in gap-variation the electrostatic stiffness is negative and  usually dominates over the mechanical, positive stiffness of the gold wire, hence the TM is unstable. Instead, in the area-variation scheme the electrostatic stiffness is much smaller than the stiffness of the wire and  the TM is stable.
In gap-variation the capacitance is inversely proportional to the gap squared, while in area-variation it is inversely proportional to the gap, but smaller gaps are known to give rise to  higher disturbances. Instead of reducing the gap, a larger surface of the electrodes is used to increase the capacity. 

In \textit{Microscope}, as a test  cylinder moves along the $X$ axis the gold wire is subjected to a longitudinal force  and responds to it with a stiffness $k_{\parallel}$; instead, in the perpendicular  $Y, Z$  directions the force is perpendicular to the wire which  responds with a different stiffness $k_\perp$.

The difference between $k_{\parallel}$ and  $k_\perp$ is well illustrated in GOCE. In  that case the  TM of each accelerometer is a parallelepiped with two ``ultra sensitive''  axes ($Y, Z$) while the third  one ($X$) is less  sensitive.  It is based on gap-variation and the gold wire ($5\,\rm \mu m$ diameter and $2.5\,\rm cm$ length)  is directed along the $X$ axis. The stiffnesses reported are $k_{\parallel}=10^{-3}\,\rm N/m$  along $X$ and $k_\perp=1.8\times10^{-6}\,\rm N/m$  along $Y,\, Z$; the gaps are of $32\,\rm\mu m$ along $X$ and  $300\,\rm\mu m$  along $Y,\, Z$, and the larger face of the parallelepiped  ($4\times1\,\rm cm^2$) is perpendicular to the $X$ axis while the $Y,\, Z$ axes are perpendicular to the smaller faces (of $1\times1\,\rm cm^2$)(\cite{E2ESimulatorGOCE}). The $X$ axis was used to test the accelerometer in the lab, where the TM must be levitated against the local gravitational acceleration instead of the much smaller tidal acceleration in orbit.

The gold wire of \textit{Microscope's}   cylinders has the same length and only a slightly bigger diameter  than the wire of  GOCE ($7\,\rm\mu m$ instead of  $5$), hence  the stiffness along the $X$ axis must be very close to  $k_{\parallel}=10^{-3}\,\rm N/m$ reported for GOCE, and indeed  the in orbit measurements found values around $\simeq10^{-3}\,\rm N/m$ for all four test cylinders (\cite{CQG2019,ChhunCQG2022-5}).

Instead,  the stiffness expected for the \textit{Microscope} wire  (\cite{BergeCQG2022-10}) was based on the theoretical formula:
\begin{equation}\label{Eq:kperp}
k_w=3\pi E\frac{r_w^4}{\ell_w^3}
\end{equation} 
with  $r_w$, $\ell_w$ the radius and length of the wire and $E=7.85\times10^{10}\,\rm N/m^2$  Young's modulus of gold,  yielding $k_w\simeq7\times10^{-6}\,\rm N/m$, about $140$ times smaller than measured.

The reason for such a  large discrepancy is that the  theoretical stiffness of  Eq.(\ref{Eq:kperp}), as well as the experimental tests supporting it (\cite{WT-RSI2000b}) apply to the response of the wire to a force acting perpendicularly to it, i.e. $k_w=k_\perp$,  whereas the relevant stiffness is that acting along the symmetry/sensitive axis, $k_\parallel$.

Within the mission team,  the  in orbit measurements by \cite{ChhunCQG2022-5} have been questioned in \cite{BergeCQG2022-10}, Sec.\,5.1. However, the model used is incorrect, giving  --for a wire along  the $X$ axis-- zero stiffness in the $Y,Z$ directions instead of the correct value resulting from Eq.(\ref{Eq:kperp}).

A higher stiffness results in a higher level of random  thermal noise from internal damping. It also affects the electrostatic control of the TM which keeps it motionless at a selected capacitive zero so that the voltage applied yields (through a scale factor in $\rm ms^{-2}/V$ established by calibration) the acceleration which caused the motion (Sec.\,\ref{Sec:LimitingNoiseAndFluctuatingZero}).

With  drag free control on, the dominant forces acting on the test cylinders are due to the Earth tides and to the mechanical stiffness
$k_{\parallel}$. 
No test cylinder is centered on the CoM of the spacecraft (Fig.\,\ref{Fig:SensorUnitsInOrbit}), hence they are subjected to a tidal acceleration. If the orbit is circular and the attitude of the spacecraft is fixed in space, the tidal acceleration along the symmetry/sensitive $X$ axis reads:
\begin{equation}\label{Eq:TidalAcceleration}
a_{tide}=\omega_{orb}^2\Big[X_{TM}\Big(\frac{3}{2}\cos{2\lambda}+\frac{1}{2}\Big)-\frac{3}{2}Z_{TM}\sin{2\lambda}  \Big]
\end{equation} 
where $\omega_{orb}=\sqrt{GM_\oplus/r^3}=2\pi\nu_{orb}$ is the orbital angular velocity, $\lambda=\omega_{orb}t$  is the instantaneous longitude relative to the spacecraft-to-Earth direction and $\vec\varrho=(X_{TM},\,Z_{TM})$ is the instantaneous position vector of the TM relative to the CoM of the spacecraft in the orbit plane.  If the orbit has a non zero eccentricity $0<e<1$, the effect is larger at pericenter than at apocenter and there is also a tidal component at $\nu_{orb}$, like the signal, although reduced by the factor $e$. If the spacecraft is spinning, tidal effects are up-converted to $2\nu_{_{EP}}$ and $\nu_{_{EP}}$. 

The difference of tidal accelerations between the two masses of each sensor unit is linear with the offsets $\Delta  X,\,\Delta Z$ between their  centers of mass within the SU. The largest effect at $2\nu_{EP}$ is then used to establish the values of these offsets, because  tidal effects are deterministic and all the parameters they depend upon are measured, and this procedure allows a better determination than  would be possible before launch. 
For SUEP the offsets measured in this way are $\Delta X\simeq20\,\mu\rm m$ and $\Delta Z\simeq-5.6\,\mu\rm m$; for SUREF (in modulus) $\Delta Z$ is similar  but $\Delta X$ is much larger ($\simeq-35.9\,\mu\rm m$) (\cite{RodriguesSystematicsCQG2022}). With SUEP measured values, the average tidal acceleration $\frac{1}{2}\omega_{orb}^2\Delta X$ as given in  Eq.(\ref{Eq:TidalAcceleration})  is $\simeq10^{-11}\,\rm ms^{-2}$. 

While the tidal effect at $2\nu_{EP}$ can be easily separated from the signal, the one at $\nu_{EP}$ competes with it. It is reduced by the very small orbital eccentricity of \textit{Microscope} ($ e\simeq10^{-4}$). In addition, as mentioned in \cite{Touboul2012}, it is almost cancelled when the spacecraft is spinning. This can be achieved by an appropriate choice of the spin frequency (it is an odd multiple of half the orbital frequency) and of the initial phase of the sensitive axis relative to the Earth so that, at pericenter and apocenter, when the tidal effect is maximum or minimum --resulting in the variation at $\nu_{_{EP}}$-- it is $90^\circ$ away from the direction to the Earth, and therefore it is not affected by the tide (Fig.\,\ref{Fig:SensorUnitsInOrbit}, right sketch).  This effect is indeed shown  negligible by \href{https://www.icra.it/mg/mg15/presentazioni/rodrigues.pdf}{M.\,Rodrigues in his talk at MGXV (15th Marcel Grossmann) meeting in Rome, 2018}. 


Under the  tidal acceleration  of  Eq.(\ref{Eq:TidalAcceleration}) and the restoring elastic force due to  $k_\parallel$ the TM will move, at any time,  to an equilibrium position  where the two effects balance each other, which occurs at an elongation $X_{eq}$ from the zero of the elastic force $O_k$. The electrostatic control acts to nullify this displacement and keep  the TM at the selected capacitive zero $O_C$. If the two zeros do not coincide, the TM in $O_C$ is subjected to a constant elastic force that the control must nullify by applying a constant voltage (bias), in addition to the sinusoidal voltage required to nullify the tidal acceleration at $2\nu_{_{EP}}$.

The resulting acceleration biases and their time evolution are reported, for all four accelerometers, in \cite{ChhunCQG2022-5}, Fig.\,9. For SUEP the bias of the outer TM is $\simeq10^{-6}\,\rm ms^{-2}$ and almost one order of magnitude smaller for the inner TM. With different offsets between their respective zeros, different masses and somewhat different stiffness, the acceleration bias is different for the two test masses of each sensor unit and the larger one dominates the acceleration difference. 
The time series  for SUEP session \#218  reported in Slide 13  \href{https://www.icra.it/mg/mg15/presentazioni/rodrigues.pdf}{of M.\,Rodrigues' talk at MGXV in Rome, 2018}, confirms that they are at the level of $10^{-6}\,\rm ms^{-2}$. For SUREF, the biases  are of a few  $10^{-7}\,\rm ms^{-2}$ (\cite{ChhunCQG2022-5}). 

For such biases orders of magnitude above the target not to impair the measurement, it is mandatory that they remain constant. This will not be the case in the presence of patch charges on the surface of the TMs  and their electrodes.
Differences in the microscopic crystal structures, and impurities, over the surface of the test masses and their surrounding electrodes give rise to local potential variations and hence to spurious forces and torques (patch effects). The smaller the gap between the surfaces, the larger the effect. Random fluctuations of these potentials translate into a $1/\sqrt{\nu}$ spectral density. Moreover, occasional dislocations of these microstructures, whatever the cause, may produce sudden variations of the mean value of the fluctuations (Sec.\,\ref{Sec:LimitingNoiseAndFluctuatingZero}).

Small force experiments require to take care of thermal effects. A well known ``killer''  in testing the WEP is the radiometer effect. Residual air pressure in the cage enclosing the TM, combined with a non zero temperature gradient along the symmetry axis, gives rise to a spurious acceleration along this axis which varies with the synodic frequency  of the spacecraft relative to the Earth, it is inevitably different for the two masses and therefore competes directly with a violation signal should the symmetry axis be  --like in \textit{Microscope}-- the sensitive axis of the WEP test.
 The radiometer effect has been a matter of concern (\cite{RadiometerPRD})  but direct in orbit measurements have shown that a very good thermal stability has been achieved  (mostly by thermal insulation of the spacecraft), such that, at the $10^{-15}$ target level of \textit{Microscope}, systematic thermal effects are smaller than  random errors (SUEP) or close to them (SUREF) (\cite{RodriguesSystematicsCQG2022}; note that the  total  residual systematic errors in Tables\,13,\,14 of this paper are  incorrect. For the correct values see Tables\,10,\,11 in \cite{TouboulCQG2022-9}.


\section{Limiting noise, patch charges  and unstable zero}\label{Sec:LimitingNoiseAndFluctuatingZero}

Ever since the preliminary \textit{Microscope} results (\cite{MS-PRL2017})  the residual random acceleration differences  in the frequency region of the signal  were attributed to thermal noise from  internal damping in the gold wires. 

In the case of internal damping the line which best fits the spectral density of the limiting noise  decreases with the frequency as $1/\sqrt{\nu}$ and reads (\cite{SaulsonInternalDamping}):
\begin{equation}\label{Eq:adiffID}
S_a^{^{1/2}}(\nu)
=f_{_{TkQ}}\cdot\frac{1}{\sqrt{2\pi\nu}}
\end{equation}
where, for  each sensor unit with TMs $m_1,m_2$, stiffnesses   $k_1,k_2$, and  the same $Q$ for both masses (if not, the lower dominates),  it holds:
\begin{equation}\label{Eq:fTkQ}
f_{_{TkQ}}=\sqrt{4k_{_B}T\cdot\left(\frac{k_1}{m_1^2}+\frac{k_2}{m_2^2}\right)\cdot\frac{1}{Q}} \ \ \ \ \ \  .
\end{equation}

For a  measurement session of duration $t_{int}$, as  long as the acceleration noise is random, and  therefore is bound to decrease as the square root of the integration time, the value  of  Eq.(\ref{Eq:adiffID}) at $\nu_{_{EP}}$  sets the limit $\delta$ to the violation signal established by the measurement session, according to the relation:
\begin{equation}\label{Eq:random}
\delta\cdot g_{_{drive}}=S_a^{^{1/2}}(\nu_{_{EP}})/\sqrt{t_{int}}
\end{equation}
where $g_{_{drive}}=7.9\,\rm ms^{-2}$ is the gravitational acceleration from Earth at the spacecraft orbital distance,  $\delta\cdot g_{drive}$ is a tiny acceleration difference between the TMs of the SU at the known signal frequency $\nu_{_{EP}}$,  and $\delta$ is  a solve-for parameter resulting from the confrontation between the predicted  acceleration differences and their measured values (as defined in \cite{TouboulCQG2022-9}, Eq.(6)).{\footnote{Note that $\delta$ is never  zero because thermal noise goes to zero only for an infinite measurement time.  Should a WEP violation signal be buried by  fluctuations around the best fitting line given by Eq.(\ref{Eq:adiffID}), the smaller their size, the shorter the additional integration  time that would be required for it to emerge and be detected. These fluctuations  yield the value $\sigma$ of the resulting $\delta\pm\sigma$ for the measurement session.}


Early on, the spin rate   was designed to be between $3$ and  $5$ times the orbital frequency $\nu_{orb}$ (\cite{Touboul2012}).  Once in orbit it was set at  $\frac{9}{2}\,\nu_{orb}$ (V2 mode).  In 2017 it was reported that at this spin rate the thermal noise of SUEP at the signal frequency of  $0.92\,\rm{mHz}$ was about $2\times10^{-10}\,\rm ms^{-2}/\sqrt{Hz}$ (Fig.\,\ref{Fig:Slide24}, top right plot), while it was expected to be 10 times lower (\cite{CQG2019}, Fig.\,11 top plot).   In order to mitigate the problem the spin rate was increased  to $\frac{35}{2}\,\nu_{orb}$ (V3 mode), moving the signal frequency  from $0.92\,\rm mHz$   to $3.11\,\rm mHz$, with an expected noise reduction by a factor  of $1.8$.
Instead, the central plot of  Fig.\,\ref{Fig:Slide24} shows that  the reduction  was by a factor of  $4$, due to the fact that in going to spin mode V3 the best fitting line given by Eq.(\ref{Eq:adiffID}) jumped downwards (together with all fluctuations around it).  At the  new signal  frequency   of  $3.11\,\rm mHz$ the limiting noise is still a factor  between  2 and 3 higher than expected   (\cite{CQG2019}, Fig.\,11 top plot).


It is apparent from Eq.(\ref{Eq:adiffID}) that the vertical position of the line (and  of the plotted curve) is dictated by the acceleration $f_{TkQ}$,  given by  Eq.(\ref{Eq:fTkQ}). The parameters that this acceleration  depends upon can hardly have changed, in going from V2 to V3 mode,  by the amount required to explain the extra gain factor of about 2.2, never mentioned in later publications and left unexplained to  this date.  
Indeed, this event may not be exceptional,  or due  to the transition from lower to higher spin rate,   since we observe very similar jumps even in sequential sessions at  the same spin rate.

\begin{figure}
\begin{center}
\includegraphics[width=0.9\textwidth]{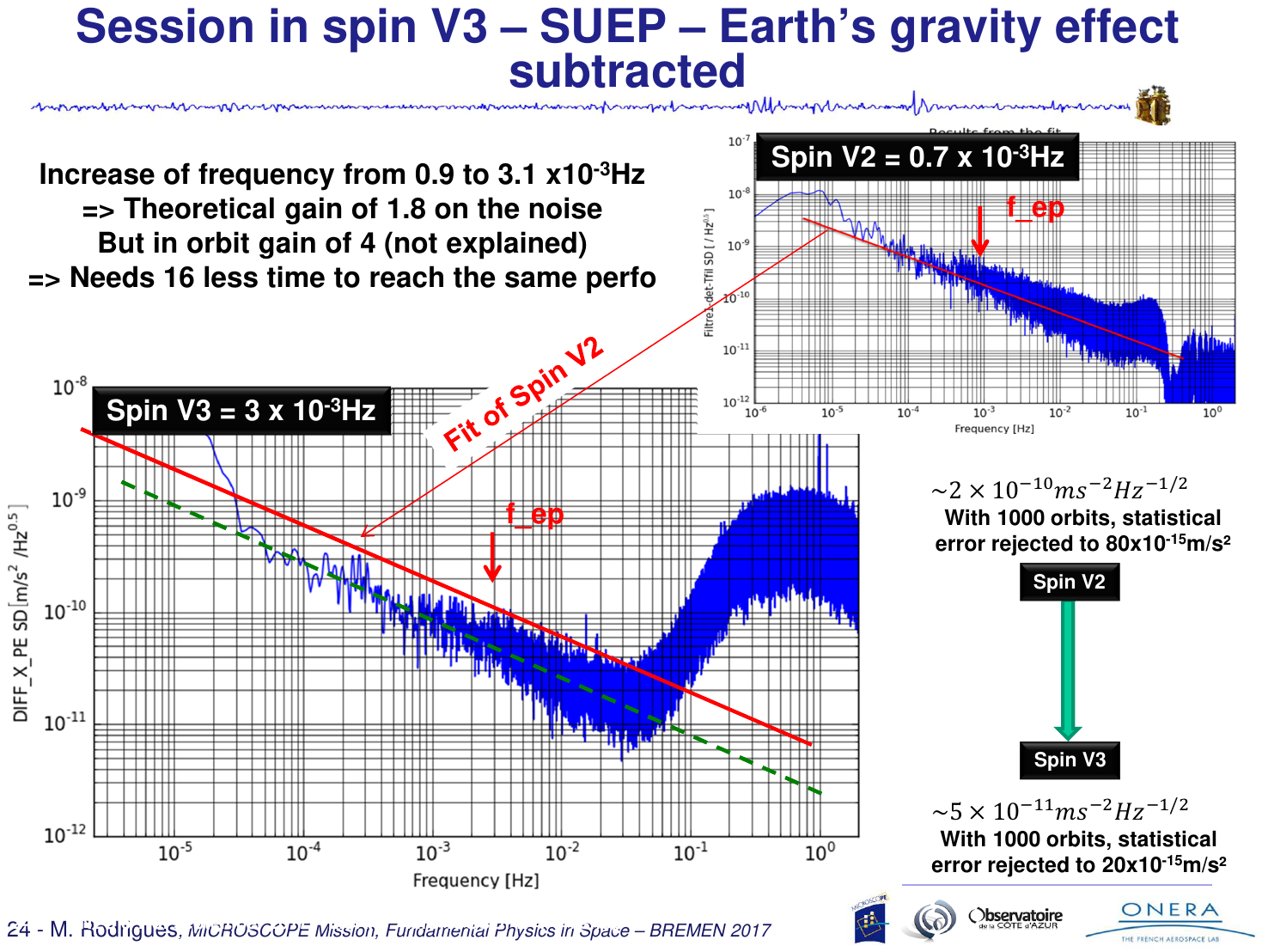}
\caption{SUEP session \#218,  Slide 24 of  the talk given by \href{https://www.zarm.uni-bremen.de/fps2017/pdf/Vortraege/Rodrigues-Microscope-breme.pdf}{M. Rodrigues at the 656th WE-Heraeus Seminar on ``Fundamental Physics in Space'' in Bremen, 2017}.   By increasing the spin frequency from mode V2 to mode V3 the frequency of the signal  $\nu_{_{EP}}$ ($f_{_{EP}}$ in the slide) is moved to a higher value where noise is lower (as $1/\sqrt{\nu}$). This will ensure a noise reduction by a factor 1.8. 
 However, it also happens that the best fitting line expressed by Eq.(\ref{Eq:adiffID}) undergoes a downward parallel translation (central plot) to a lower level of noise at all frequencies, including the new   $\nu_{_{EP}}=3.11\,\rm mHz$, such that the total noise reduction is by a factor of 4 (unexplained).
}
\label{Fig:Slide24}
\end{center}
\end{figure}

As the key limiting factor of each measurement, and of the final \textit{Microscope} test of WEP,  the quantity $S_{a}^{^{1/2}}(\nu_{_{EP}})=f_{TkQ}/\sqrt{2\pi\nu_{_{EP}}}$ deserves a careful scrutiny. 
For a given value of $f_{_{TkQ}}$  the measured   noise is fitted by the same straight line.  Once the oscillator has been set up and launched, $k$'s and  $Q$'s are fixed, and (except for a mild dependence on $T$) the value of $f_{_{TkQ}}$ is fixed and so is the straight line of Eq.(\ref{Eq:adiffID}) and its value at $\nu_{_{EP}}$. 

We cannot exclude a variation of the stiffness after a calibration, because in the  calibration the quadratic acceleration term must be kept sufficiently small, which is achieved by offsetting the TM to a new zero, hence affecting the stiffness,  $f_{_{TkQ}}$ and  $S_a^{^{1/2}}(\nu_{_{EP}})$ (G.\,Catastini, private communication). Conversely, in sequential sessions, or within the same session analyzed with equally valid methods, we expect no such variations.  
Should any anomalies occur, they will clearly appear in the values of $Q$ because $Q\propto1/S_a(\nu_{_{EP}})$. Moreover, the extent and the relevance of the anomaly can be assessed by comparison with   $Q$ measurements carried out in ground tests of WEP as well as in experiments for the measurement of the universal constant of gravity  and the detection of gravitational waves.

In order to carry out this analysis we need the values of $S_a^{^{1/2}}(\nu_{_{EP}})$ for all the valid sessions, but there is no table listing these values in the \textit{Microscope} publications, as there is for  $\delta_i\pm\sigma_i$ (\cite{TouboulCQG2022-9}) although  accelerations are the measured physical quantities while $\delta$ is a derived number (see \cite{TouboulCQG2022-9}, Eq.\,(6)) and moreover
 $S_a^{^{1/2}}(\nu_{_{EP}})$   does not depend on the integration time --unlike $\delta$.
With  all $\delta$  values from \cite{TouboulCQG2022-9}, Tables 6,\,7    and the duration of each session  from \cite{MissionScenarioCQG2022} we   use Eq.(\ref{Eq:random}) to derive all the corresponding values of   $S_a^{^{1/2}}(\nu_{_{EP}})$. 
The correctness of our derived values can be  checked when the measured ones  are published.{{\footnote{In\,\cite{ChhunCQG2022-5}, Fig.\,11, the values of  $S_{a}^{^{1/2}}(\nu_{_{EP}})$ are plotted  for all sessions, including those which have not contributed to the final result of the WEP test, the purpose being to demonstrate an overall stability after the first few sessions of the mission. Although the numerical values of the plotted marks are not easy to read in the vertical scale used,  for the 120-orbit SUEP session \#236 (V3 mode, blue curve in Fig.\,11, error bar shown on V2  mark by mistake) we can see that the plotted mark  is much too high (by at least a factor of 3) with respect to the best fit value of about  $2\times10^{-11}\,\rm ms^{-2}/\sqrt{Hz}$ reported in Fig.\,6, \cite{TouboulCQG2022-9}.  The discrepancy is confirmed 
 using Eq.(\ref{Eq:random}), because  the plotted mark leads to a value of $\delta_{236}$ at least 3 times higher than reported in \cite{TouboulCQG2022-9}, Table\,7.
For other sessions of 120-orbit duration like \#236, such as \#238, \#254 and \#404 we know from Eq.(\ref{Eq:random}) that the ratios of their  $S_{a}^{^{1/2}}(\nu_{_{EP}})$ values to that of session \#236 must be the same as the ratios of the corresponding values of $\delta$  (available in Table\,7 of \cite{TouboulCQG2022-9}). While  these ratios are never close to 1,  their individual marks are all plotted at about the same height  in Fig.\,11 of \cite{ChhunCQG2022-5}.  Thus, if  this figure is correct, it contradicts Table\,7 of  \cite{TouboulCQG2022-9} as well as Fig.\,1 of \cite{MS-PRL2022} and the final \textit{Microscope} test of WEP published therein.
}}

Then, given the equilibrium temperature (\cite{TouboulCQG2022-9}), the masses and the measured stiffnesses,  the quality factor $Q$ is inferred from Eq.(\ref{Eq:fTkQ}). Since $k_1\, ,k_2$ may vary because of calibrations we compute also $k/Q$ assuming the same unknown $k$ for both TM's. These quantities are listed in Table\,\ref{table:SUEPnew} for SUEP and in Table\,\ref{table:SUREFnew} for SUREF, along with the percentage  $p$ of artificial data that have been introduced in each session  after the elimination  of glitches. 

Tables\,\ref{table:SUEPnew} and\,\ref{table:SUREFnew} show that large jumps in $S_{a}^{^{1/2}}(\nu_{_{EP}})$, $Q$ and $k/Q$ occur between sequential sessions  (numbered by even numbers) in three  cases for SUEP and  SUREF.

\begin{table} [ht] \tiny\vspace{-2.8mm}
\caption{SUEP Sensor Unit. Session number in V2 or V3 spin mode; session $\delta$ and  percentage of glitches $p$ from Tables 6,7 and 4,5 of \,\cite{TouboulCQG2022-9}; calculated values of  $S_{a}^{^{1/2}}(\nu_{_{EP}})$, $Q$, $k/Q$. $\downarrow$ indicates  a jump between  sequential sessions; * an  $M$ vs $A$ discrepancy (see text below).} 
\begin{tabular}{|>{\raggedright\arraybackslash}m{1.1cm}| 
>{\raggedleft\arraybackslash}m{0.6cm}| 
>{\raggedleft\arraybackslash}m{0.6cm}| 
>{\raggedleft\arraybackslash}m{0.8cm}| 
>{\raggedleft\arraybackslash}m{0.8cm}| 
>{\raggedleft\arraybackslash}m{0.5cm}| 
>{\raggedleft\arraybackslash}m{0.5cm}| 
>{\raggedleft\arraybackslash}m{0.7cm}| 
>{\raggedleft\arraybackslash}m{0.7cm}| 
>{\raggedleft\arraybackslash}m{0.4cm}|}
\hline
 &  
 \multicolumn{2}{| c |}{$\delta$} & 
 \multicolumn{2}{| c |}{$S_{a}^{^{1/2}}(\nu_{_{EP}})$} & 
 \multicolumn{2}{| c |}{$Q$} &
   \multicolumn{2}{| c |}{$k/Q$} &
 p
 \\
 
 &  
 \multicolumn{2}{| c |}{$10^{-15}$} &  
 \multicolumn{2}{| c |}{$10^{-11}\rm ms^{-2}/\sqrt{Hz}$} &
 \multicolumn{2}{| c |}{} &
  \multicolumn{2}{| c |}{$10^{-3}\rm{N/m}$} &
 \%   
 \\ \hline
 
Session&
$M$ &
$A$  &
$M$ &
$A$ &
$M$ &
$A$  &
$M$&
$A$&
 \\
\hline

{$\downarrow$210\,(V3)}& ${-30.1}$ &${-29.2}$ & {13 }&{13}  & {0.75}&0.79 & 1.2&1.1 &{18} \\ \hline 

{212\,(V3)} & ${10.4}$ & ${9.5}$ & {4.9} &{4.5}    &{5.2}&{6.3} &0.17 & 0.15 &{17}  \\ \hline\hline

    *218\,(V3) & $3.6$ & $6.7$ & 2.4 &  4.5 &  22& 6.3 &0.042 &0.14  &15  \\ \hline\hline

  234\,(V3) & $5.6$ & $5.9$ & 3.3 & 3.4  &12&11 & 0.078& 0.086  &18  \\ \hline

  $\downarrow$236\,(V3) & $2.7$ & $2.6$ &  1.8 & 1.7 &39& 42 &0.024 &0.022 &21  \\ \hline 

  238\,(V3)  & $6.1$ & $5.8$ & 4.1 & 3.9 &7.6&8.4 &0.12 & 0.11 &24  \\ \hline\hline

  252\,(V3) & $-14.7$  & $-14.9$ & 9.2 &9.3 &1.5&1.4  &0.62 &0.64 &26  \\ \hline 

 $\downarrow$254\,(V3) & $-14.2$ & $-14.1$ & 9.5 & 9.4   &1.4& 1.4 & 0.65& 0.64 &27 \\ \hline 

  256\,(V3) & $-4.7$ & $-5.3$  & 3.1 &3.5    &13&10 &0.071 & 0.091 &28  \\ \hline\hline

    326-1\,(V3)& $-10.1$ & $-16.3$ &  4.8&  7.7  &   5.5&  2.1 &0.16 &0.43 &12 \\ \hline

  326-2\,(V3) & $-11.1$  & $-10.4$ & 3.9 &  3.7  &8.1 & 9.2  &0.11 & 0.099&7 \\ \hline\hline

    358\,(V3) & $15.4$ & $15.8$  & 9.0 & 9.2   &  1.6&   1.5 &0.59 & 0.62 &14 \\ \hline \hline

  402\,(V2) & $27.3$  & $28.4$ & 7.1 & 7.3   &8.5& 7.9  &0.11 & 0.12&35 \\ \hline \hline

   *404\,(V3) & $6.3$ & $4.7$ & 4.2 & 3.1   & 7.1& 13 &0.13 & 0.071&23   \\ \hline 

  406\,(V3) & $6.0$ & $5.9$ &  1.6& 1.6  & 47&49 &0.019 & 0.019 &23    \\ \hline\hline

    *438\,(V2) & $-12.5$  & $-23.4$ & 4.3 &  8.1  & 23&  6.5 & 0.040& 0.14 &21  \\ \hline\hline

     *442\,(V2) & $-10.7$  & $-1.5$ & 4.1 &0.58    &    25&   1273 &0.037 &0.00072  &21 \\ \hline\hline

    *748\,(V2) & $-17.5$ & $-23.4$ & 5.2 & 7.0   &15&8.7  & 0.059& 0.11&25 \\ \hline \hline
  \hline%

    750\,(V3) & $66.6$ & $66.9$ & 11 & 12   &  0.97&  0.95  & 0.94&0.97 &19  \\ \hline 
    \end{tabular}
\label{table:SUEPnew}
\end{table}

It is interesting that  for SUEP the jumps in  $S_{a}^{^{1/2}}(\nu_{_{EP}})$ are by a factor between 2 and 3, close to the one  reported  in Fig.\,\ref{Fig:Slide24} in the transition  from spin mode V2 to V3. However, all three jumps reported in Table\,\ref{table:SUEPnew} involve sessions in V3 mode only, indicating  that they  are unrelated to a variation of the spin rate.

For SUREF the three largest jumps are even larger, and in this case the sessions involved are all in V2 mode.

Values  of $Q<1$ and close to critical damping observed in two SUEP  sessions are unrealistic.

Two methods of analysis, $M$-$ECM$ ($M$) and $ADAM$ ($A$), have been employed to estimate the experiment parameters, including  $\delta$ and $\sigma$. Both methods use artificially reconstructed data to account for the missing data resulting from  the elimination of glitches.\footnote{Method $M$ estimates in the time domain the missing data,  maximizing the likelihood conditional on the observed data, and then computes the least-squares estimate of the regression parameters. An estimation of the  SD is also produced in the process. Method  $A$ performs the parameter estimation in the frequency domain. As such it requires an uninterrupted, regularly spaced time series, which is obtained by filling the gaps left by the removal of glitches with the artificial data estimated by $M$.
The two methods are stated to be equivalent, but in a few cases the values of $\delta$
calculated by $M$ and $A$ differ considerably (by a factor $\simeq2$ in sessions \#218, \#438 and by a factor 7 in \#442),
implying that the same data stream leads to a different SD according to one or the other method.
%
SUEP session  \#442 is an instructive extreme case.  As  reported in\,\cite{TouboulCQG2022-9}, Table\,7,  
the noise is essentially the same but the values of $\delta$ differ by a factor 7; $\delta$ representing a level of WEP violation compatible with the measured acceleration difference noise. That is, for the same measurement session, in the same experimental conditions,  the same data with the same level of noise lead  to almost one order of magnitude difference in the evaluation  of  $\delta$.
The difference is particularly evident when the results are interpreted in terms of the physics of the dominant thermal noise from internal damping (widely different values of $Q$, up to an unrealistic large value in session  \#442).
SUREF session \#778-1  is the only one, among all 32  sessions ($19$ SUEP $+$ $13$ SUREF), not affected by glitches, hence without artificial data. In this case,  the results are exactly the same for $\delta$ and only slightly different for $\sigma$, as one expects if the same data set is analyzed with two equally valid methods (\cite{TouboulCQG2022-9}, Table 6).
Sessions  \#442 and \#778-1 show  that the $M$ and $A$ methods are in agreement when applied to a time series of real measured data, but give different results (in particular different $\delta$ values) in the presence of artificial reconstructed data.  Since  the artificial data in  $A$ are those estimated by $M$, it is the way the two methods manipulate these data that makes the difference.}

\begin{table} [ht] \tiny\vspace{-2.8mm} 
\caption{Same as Table\,\ref{table:SUEPnew} for SUREF Sensor Unit. A $\bullet$ has been added to indicate sessions with $Q$ values much larger than the largest value of $118$ measured in ground tests (\cite{WT-RSI2000b}).}
\begin{tabular}{|>{\raggedright\arraybackslash}m{1.5cm}| 
>{\raggedleft\arraybackslash}m{0.57cm}| 
>{\raggedleft\arraybackslash}m{0.57cm}| 
>{\raggedleft\arraybackslash}m{0.8cm}| 
>{\raggedleft\arraybackslash}m{0.8cm}| 
>{\raggedleft\arraybackslash}m{0.45cm}| 
>{\raggedleft\arraybackslash}m{0.45cm}| 
>{\raggedleft\arraybackslash}m{0.7cm}| 
>{\raggedleft\arraybackslash}m{0.7cm}| 
>{\raggedleft\arraybackslash}m{0.3cm}|}
\hline
 &  
 \multicolumn{2}{| c |}{$\delta$} & 
 \multicolumn{2}{| c |}{$S_{a}^{^{1/2}}(\nu_{_{EP}})$} & 
 \multicolumn{2}{| c |}{$Q$} &
   \multicolumn{2}{| c |}{$k/Q$} &
 p 
 \\
 
 &  
 \multicolumn{2}{| c |}{$10^{-15}$} &  
 \multicolumn{2}{| c |}{$10^{-11}\rm ms^{-2}/\sqrt{Hz}$} &
 \multicolumn{2}{| c |}{} &
  \multicolumn{2}{| c |}{$10^{-3}\rm{N/m}$} &
 \%   
 \\ \hline
 
Session&
$M$ &
$A$  &
$M$ &
$A$ &
$M$ &
$A$  &
$M$&
$A$&
 \\
\hline

     $\downarrow$*$\bullet$120-1\,(V2) & $-3.1$ & $-4.2$ &0.89 &1.2  & 268& 146 &0.042 & 0.077&4 \\ \hline 
  120-2\,(V2) & $-16.8$  & $-15.1$  &8.2 &7.4    &3.1&3.9 & 0.36& 0.29&15  \\ \hline\hline
   $\downarrow$174\,(V2) & $7.8$  & $8.0$ &4.4 &  4.5 &11&10&0.10 &0.11 &25 \\ \hline 
   $\bullet$176\,(V2) & $1.7$  & $1.8$  & 0.82 & 0.86  & 317& 283&0.0036 &0.0040 &40  \\ \hline\hline
   294\,(V3)  & $-8.0$  & $-7.7$ &4.2 & 4.1 & 3.5& 3.7&0.33 & 0.30&17 \\ \hline\hline
376-1\,(V2)  & $-3.4$  & $-4.1$  & 1.2 & 1.5 &135& 93&0.0083 &0.012 &14\\ \hline 
 376-2\,(V2) & $-5.7$ & $-6.4$ & 1.8 &2.1 &62& 49&0.018 &0.023 &11  \\ \hline \hline
  380-1\,(V3) & $7.6$ & $7.4$ &3.1 & 3.1   &6.4&6.7 &0.18 & 0.17&7   \\ \hline 
  380-2\,(V3) & $9.3$  & $8.9$  &3.3 &3.2    &5.7& 6.3&0.20& 0.18&5 \\ \hline \hline
452\,(V2)  & $-4.3$  & $-4.8$  & 1.5&  1.7  &101& 81& 0.011& 0.014&20 \\ \hline 
 454\,(V2) & $-3.1$  & $-3.7$   &1.4 &  1.7  & 111&78 &0.010 & 0.014&22 \\ \hline\hline
   $\downarrow$778-1\,(V2) & $-8.1$  & $-8.1$  &3.0& 3.0   &23&23& 0.049&0.049 &0 \\ \hline 
  *$\bullet$ 778-2\,(V2) & $-2.3$  & $-3.2$  &0.59 & 0.83   & 599&  309&0.0019 &0.0036 &6  \\ \hline 
    \end{tabular}
\label{table:SUREFnew}
\end{table}

In SUEP session \#218 the  values of  $\delta$ differ by a factor of about 2 depending on
the type of analysis. This session (120 orbits) was the basis of the early results (\cite{MS-PRL2017}) and was further
elaborated two years later (\cite{CQG2019}), ending up with a different value of the spectral density at the signal frequency.
Even discarding the earlier results, it is apparent that the latest analysis cannot be considered conclusive. $Q$ values
differing by a factor 3.5, for the same oscillator in the same conditions, are
inexplicable. 

The reference sensor SUREF behaves quite differently from SUEP.   Thermal noise is about a factor $\sqrt{2}$ smaller than for SUEP  (assuming the same $Q$), due to a factor $2$ smaller  $(k_1/m_1^2+k_2/m_2^2)$ term,  and residual systematic errors are comparable to  the stochastic errors (\cite{TouboulCQG2022-9}). Four  out of nine sessions have been split  because of sudden jumps in the mean value of the differential acceleration  that required the two data segments to be treated as distinct experiments (in SUEP only one session, \#326,  out of 18 has been split).
 Unlike glitches, these jumps do not occur on all accelerometers simultaneously, hence they cannot be attributed to the spacecraft but are likely to originate in the accelerometers themselves, and last much longer (tens of seconds). 
In SUREF, in addition to the three cases   of anomalous jumps mentioned above, we observe three sessions with large $M$ vs $A$ discrepancies (see Table\,\ref{table:SUREFnew}). The values of $Q$ are typically larger than in SUEP and in three cases much larger than the largest value of $118$ reported in ground measurements (\cite{WT-RSI2000b}).

Early results (\cite{MS-PRL2017})  were based for SUREF on session \#176 (62 orbits). The spectral density  $S_{a}^{^{1/2}}(\nu_{_{EP}})$ reported in\,\cite{MS-PRL2017} was confirmed in\,\cite{CQG2019}  and attributed to $1/\sqrt{\nu}$ thermal noise,  yielding  $\delta=3.75\times10^{-15}$ and  $Q=65$, consistent with ground measurements. Instead, the final values  of $\delta_{{M}}$ and  $\delta_{{A}}$  are much smaller (Table\,\ref{table:SUREFnew}), and the  corresponding  $Q$'s  of  317 and 283  are far too high to be realistic on the basis of ground tests. Note that session \#176 has $40\%$ of reconstructed data.

A question naturally arises: is there another source of noise with the same $1/\sqrt{\nu}$ dependence like thermal noise from internal damping which may explain the  anomalies described above?

As shown in Sec.\,\ref{Sec:ForcesControl}, the control of the TMs requires to control large acceleration biases.  The acceleration bias $a_{bias}$ and the voltage  $V_{bias}$ applied to control  it are related through the scale factor $s_{_f}$:
\begin{equation}\label{Eq:AccelerationVoltageBias}
a_{bias}=s_{_f}V_{bias}
\end{equation}
which, in the area-variation scheme used for the sensitive axis, reads:
\begin{equation}\label{Eq:ScaleFactor}
s_{_f}=\frac{2\pi\epsilon_\circ(2r)}{md}\cdot(V_p-V_p')
\end{equation}
where $\epsilon_\circ$ is the dielectric constant of vacuum, $m$ the mass of the test cylinder, $r$ the radius of the electrode belt, $d$ the gap, $V_p$ the polarization voltage of the cylinders  and $V_p'$ the voltage applied to the electrodes ($5\,\rm V$ and $-2.5\,\rm V$  respectively). The presence of patches of charges on the  surface of the test cylinder and  the electrodes would generate a  patch potential in addition to $V_p$ and $V_p'$, and patch potentials are known to undergo low frequency variations. An additional fluctuating patch potential $\delta V_{fpp}$ gives rise to an additional acceleration $\delta a{_{fpp}}$ proportional to the product of this spurious potential times the bias control potential:
\begin{equation}\label{Eq:PatchAcceleration}
\delta a{_{fpp}}=\frac{2\pi\epsilon_\circ(2r)}{md}\cdot \delta V_{fpp}\cdot V_{bias}\ \ \ \  .     
\end{equation}
Using Eqs.(\ref{Eq:AccelerationVoltageBias}) and (\ref{Eq:ScaleFactor}), it leads to the relation
\begin{equation}\label{Eq:Ratio}
\frac{\delta a{_{fpp}}}{a_{bias}}=\frac{\delta V_{fpp}}{V_p-V_p'}    
\end{equation}
showing that in the presence of  a large acceleration bias even a very small fluctuating patch potential can make the zero of the measurement fluctuate and result in a non negligible fluctuating patch acceleration.

As recalled in Sec.\,\ref{Sec:ForcesControl}, the difference of acceleration biases between the TMs of SUEP is  $\simeq10^{-6}\,\rm ms^{-2}$.   According to Eq.(\ref{Eq:Ratio}),  a  fluctuating patch potential $\delta V_{fpp}\simeq0.3\,\mu V$ at the V3 signal frequency $\nu_{_{EP}}=3.11\,\rm mHz$ would produce a spurious acceleration   of $4\times10^{-14}\,\rm ms^{-2}$, as large as  the average acceleration difference measured in all SUEP sessions (once normalized by session duration) at $\nu_{_{EP}}$ (\cite{ChhunCQG2022-5}, Fig.\,10) and enough to change it by $100\%$ either way.

In GOCE a source of low frequency acceleration noise with $1/\sqrt{\nu}$ dependence below $0.01\,\rm Hz$ was found to be due to  fluctuating potentials generated  by charge patches  on the test mass and its electrodes (\cite{E2ESimulatorGOCE}). GOCE was set up as a gravity gradiometer with $0.5\,\rm m$ separation of the TMs and a measurement bandwidth from $500\,\rm mHz$ to $0.1\,\rm Hz$. At $3\,\rm mHz$ the  acceleration  difference sensitivity  was of  $2\times10^{-12}\,\rm ms^{-2}/\sqrt{Hz}$ along the ultra sensitive $Y, Z$ axes  (more than one order of magnitude better than that of  SUEP sensitive axis)  and $4\times10^{-10}\,\rm ms^{-2}/\sqrt{Hz}$ along the least sensitive $X$ axis, with a contribution from patch fluctuating potentials of  about  $4\times10^{-13}\,\rm ms^{-2}/\sqrt{Hz}$ and $2\times10^{-10}\,\rm ms^{-2}/\sqrt{Hz}$ respectively (\cite{TouboulReview2016}).\footnote{With no wire  the zero of each TM  control  (dictated by geometry, construction, mounting and  electronics) does not coincide with the zero of other forces acting  on the test mass, such as Earth tides and electrostatic stiffness. Applying a constant bias is therefore inevitable in the \textit{Microscope} design, regardless of the gold wire. Hence, any fluctuation of the bias in the frequency region of the WEP violation signal is relevant when aiming at very high precision.}

As a measurement session is completed and a new one is initiated, patches of electric charges rearrange themselves,  giving rise to new fluctuations. Hence, their contribution to the acceleration noise changes, either upwards or downwards, which may cause jumps  like the observed ones --unexplained so far.

In \textit{Microscope} the requirement on patch potentials is that they shall not exceed $15\,\rm mV$ DC. Our analysis shows that the issue of their variations at low frequencies should be revisited,  based on knowledge --both theoretical and experimental-- acquired with GOCE, because it may give rise to an unstable zero.

An unstable zero is  a known issue in WEP tests, related to the problem of initial conditions (or release) errors,  affecting all tests  not  designed as intrinsically null experiments, including those with laser tracked satellites, celestial bodies and cold atoms (\cite{Blaser2001,GRG-2008,Nobili2016}). In other tests, primarily those based on torsion balances,  the physical observable is a true null (\cite{PerBraginskyPLA2018}).

\section{Outliers,  gaps and artificial data}\label{Sec:OutliersGapsArtificialData}

The experiment was plagued by anomalous short-duration ($<5\rm s$) acceleration spikes originating in the
spacecraft and occurring simultaneously in the four test masses.

Since 2001 five space missions carrying ONERA’s accelerometers have been launched   with a total of seven satellites, and all but GOCE  have  experienced such spikes. There is wide consensus that the spikes are triggered by  energy inputs from the Earth  and the Sun causing micro-vibration, e.g.  release of stress energy in the spacecraft  multi-layer insulation (MLI). At the time of the design of GOCE, reports of such effects  caused alarm, and countermeasures were adopted at design and test level based on a stiffer insulation, which were successful, and no spikes were seen.
 
The MLI  is subjected to  temperature variation cycles during the  motion of the  spacecraft relative to the Earth and the Sun, which act as sources of heat in contrast to cold deep space. If MLI  is not rigidly attached, stress energy accumulates on some spots   at the synodic frequencies relative to the sources of energy inputs. The release events of the  accumulated energy are erratic, as they depend on unpredictable factors, but  they tend to occur at the synodic frequencies and their harmonics.
\begin{table} [ht] \vspace{-3mm}
\caption{Synodic frequencies of the \textit{Microscope} spacecraft  relative to the Sun, $\nu_{{syn\odot}}$ (left column; fewer digits are shown because it is approximated by the spin frequency) and the Earth, $\nu_{_{syn\oplus}}$ (right column) up to the 5th harmonic. All frequencies are given in $\rm mHz$ and refer to spin mode V3 ($\nu_{_{spin}}=N\nu_{_{orb}}/2$ with $N=35$).} 
\vspace{3mm}
\begin{tabular}{|>{\raggedright\arraybackslash}m{4cm}|  
>{\raggedleft\arraybackslash}m{4cm}|} 
\hline
 \multicolumn{2}{| c |}{$\nu_{_{orb}}=0.16818\,\rm mHz$}
 \\
 \multicolumn{2}{| c |}{$N=35$}    
 \\ \hline\hline
 
$\nu_{{syn\odot}}\simeq\nu_{_{spin}}=N\nu_{_{orb}}/2$ &
$\nu_{_{syn\oplus}}=\nu_{_{orb}}+\nu_{_{spin}}=(N+2)\nu_{_{orb}}/2$ 
 \\
\hline \hline\hline
  $N\nu_{_{orb}}/2$ & $(N+2)\nu_{_{orb}}/2$\\ 
  $2.943$  & $3.11133$    \\ \hline\hline
   $2N\nu_{_{orb}}/2$ & $2(N+2)\nu_{_{orb}}/2$ \\ 
   $5.886$  & $6.22266$   \\ \hline\hline
   $3N\nu_{_{orb}}/2$  & $3(N+2)\nu_{_{orb}}/2$  \\ 
   $8.829$  & $9.33399$  \\ \hline \hline
   $4N\nu_{_{orb}}/2$ &  $4(N+2)\nu_{_{orb}}/2$    \\  
   $11.772$  & $12.44532$ \\ \hline\hline 
   $5N\nu_{_{orb}}/2$  &  $5(N+2)\nu_{_{orb}}/2$  \\ 
   $14.715$  & $15.55567$  \\ \hline\hline 
    \end{tabular}
\label{table:SpikeFrequencies}
\end{table}

In \textit{Microscope} the transfer function of the two test cylinders in the sensor unit are not the same; in addition, the force is non gravitational and the masses are different. As a result, spikes are observed in the acceleration differences of each SU.

\textit{Microscope's }synodic frequency relative to the Sun is well approximated by the spin frequency relative to  the fixed stars, the difference being the annual orbital frequency of the Earth around the Sun of $2\times10^{-7}\,\rm rad/s$. The synodic frequency relative to the Earth is  the orbital plus the spin frequency, i.e. the frequency of the signal ($\nu_{_{EP}}=3.11133\,\rm mHz$ in V3 mode). 

Table\,\ref{table:SpikeFrequencies} lists the frequencies at which the effects of glitches are likely to occur, and we  may expect to see an effect  also at  the difference between these two frequencies, i.e. the orbital frequency.

An FFT analysis of  large glitches (with $\rm SNR>3$)  for a typical measurement session is reported in Fig.\,7 of \,\cite{MicroscopeGlitchesCQG2022-008}.
 It shows that the large glitches occur at all the frequencies  listed in Table\,\ref{table:SpikeFrequencies} at the level of $10^{-11}\,\rm ms^{-2}$ and above, reaching about $7\times10^{-11}\,\rm ms^{-2}$ at the spin frequency, i.e. the synodic frequency relative to the Sun. Higher harmonics appear too, but they are less relevant to the experiment. 
In the logarithmic scale of the Figure each line appears as a double line because the two synodic frequencies are close to each other (see Table\,\ref{table:SpikeFrequencies}),  except the line at the orbital frequency which is the difference of the two. However, in  the inset of the same Figure showing a zoom in the region of  the spin frequency, the scale is linear and the  two synodic frequencies  clearly appear as distinct lines.

In\,\cite{MicroscopeGlitchesCQG2022-008} it is concluded that glitches  must be removed from the data because it is impossible to accurately model their effects which include one at the  frequency of the signal.
In the time series of each session glitches are identified  as all outliers above $4.5\,\sigma$ from the moving average, some time is allowed  after each outlier to account for  its decay,  and all data points identified in this way are removed, up to 35\% in SUEP and up to 40\% in SUREF. 
As a result, the time series of the remaining data is interspersed with gaps. Gaps are then filled with artificial reconstructed data so that the resulting time series can be analyzed in the frequency domain to compute the spectral density of each measurement session, hence the limit to WEP violation $\delta_i\pm\sigma_i$. By including all $19$  SUEP  sessions the final limit $\eta=\delta_{fin}\pm\sigma_{fin}=(-1.5\pm2.3)\times10^{-15}$ reported in \cite{MS-PRL2022} is obtained,  with: 
\begin{equation}\label{Eq:deltafin}\vspace{-3mm}
\delta_{fin}=\bigg({\sum\limits_{i=1}^{i=19}\frac{\delta_i} {\sigma_i^2}}\bigg)\sigma_{fin}^2 \end{equation}
and $\sigma_{fin}^2 =1/\big({\sum\limits_{i=1}^{i=19}\frac{1}{\sigma_i^2}}\big)$ (see \cite{TouboulCQG2022-9}).

This way of dealing with outliers raises two issues. 

The first one is related to the fact that gaps retain memory of the glitches whose elimination originated them; and since the sign of WEP violation is not known, they may in fact mimic a violation signal or cancel an effect, signal or systematic.  

The second issue is that the artificial data at the level of  random platform noise which replace the outliers do contribute to increasing the integration time of each measurement session, hence to reducing the level $\delta_i$ of WEP violation of the session (Eq.\ref{Eq:random}), i.e. to improving the test. Then, according to  Eq.(\ref{Eq:deltafin}), the final result of the whole experiment is also improved. If this procedure were correct, the more outliers are present, the larger the fraction of gaps, the larger the fraction of artificial data, the better the WEP test of the session and the WEP test of the entire experiment. Which is clearly a paradox. No matter how artificial data  are generated and/or manipulated, they should not  introduce any  physical information, i.e. their effect on the $\delta$ estimated in each session ought to be null.


An interesting  comparison is reported in\,\cite{MicroscopeGlitchesCQG2022-008} between the FFT of the differential acceleration obtained from the original measured data, and the FFT    after glitches were removed and the missing data were reconstructed. This was done for  SUREF session \#380 lasting 120 orbits. This session has been chosen because, with equal composition test masses, any effect at $\nu_{_{EP}}$ is certainly spurious and moreover the low  noise makes the effects of glitches well visible. 
 %
%
Fig.\,10 of \cite{MicroscopeGlitchesCQG2022-008} shows the two plots, FFT of the measured data in black, of reconstructed data in red. All the double lines listed in Table\,\ref{table:SpikeFrequencies} (plus the single line at $\nu_{_{orb}}$) are clearly visible (the tidal line at $2\nu_{_{EP}}$ has not been removed). It is apparent that the effects of glitches, including the one  at  $\nu_{_{EP}}$, have been reduced but not eliminated. In fact, in the final SUREF analysis this session, originally with $46\,\%$ of artificial  data, has  been split into two short sessions (of $46$ and $34$ orbits), each one with a small percentage of glitches ($7\,\%$ and $5\,\%$). Thus, most of the reconstructed data have been eliminated, which is a reasonable  decision. 


Since the  artificial data have an impact also at the frequency of the signal a check has been performed to 
demonstrate the correctness of the procedure. 

After establishing the values of   $\delta_i$ for all the measurement sessions (\cite{TouboulCQG2022-9}, Table\,6  for SUREF and Table\,7 for SUEP),  the  raw data of the original time series of each session were taken and, before processing,  a fake signal was introduced in each of them, first at the   \eotvos\ level of $3.4\times10^{-15}$ ($2.7\times10^{-14}\,\rm ms^{-2}$ acceleration difference) and then at one order of magnitude bigger. Each time series including the fake signal was processed with the same procedure used to generate the original $\delta_i$,  thus generating  two new series of $\delta_i$  which should differ from the original one by the fake signal only (\cite{TouboulCQG2022-9}, Tables\,8 and\,9).
 If  for each session the difference yields the corresponding fake signal, this is taken as evidence that the procedure is correct because, should the time series of the raw data contain a real violation signal at these levels, it would be recovered correctly.
 
Yet, in all cases $\delta_i$ has been obtained with the same procedure, eliminating the same outliers, generating the same gaps, filling the gaps with the same data, the difference being only the  fake signal which is therefore obviously recovered. The problem  is that   $\delta_i$ itself may  be affected by the artificial data, and this fact is not addressed by the check.

A clear-cut  way out is to proceed in the footsteps of the \textit{E\"ot-Wash} ground tests of the WEP with rotating torsion balances.  They established the limit to WEP violation using only the real measured data by exploiting  the fact that the frequency and phase of a differential acceleration due to WEP violation in the field of a source body (the Earth in this case) are known, only its amplitude and sign are unknown. 

In the case of\textit{ Microscope} it works as follows. At any given time, in the reference frame rotating with the spacecraft, the position  of the Earth and the phase of the sensitive axis are known. Hence, in the time series of the acceleration differences  the violation signal is a sinusoid with the synodic frequency  of the spacecraft relative to the Earth whose maximum size (of opposite sign and unknown modulus) occurs twice per synodic period,  when the sensitive axis points towards or away  from Earth  (sign unknown).  This demodulated phase lock-in signal can be fitted to the time series of the acceleration differences --only  the real measured ones-- in order to determine its amplitude and sign, which therefore would not be affected by whatever gaps.

A problem is that in \textit{Microscope} outliers  have an  effect at the  frequency of the signal. However, we do not expect it to have the same phase. The differential acceleration of a potential WEP violation in the field of the Earth is a vector pointing to the center of mass of the Earth, hence it is maximum (in amplitude) when the coaxial  cylinders of  the sensor unit are along the radial direction. We have seen that glitches induced by heat from the Earth  produce an acceleration difference at the same frequency, but they are certainly unrelated to the phase of the symmetry/sensitive axes (when the accumulated stress energy is released the test cylinders may be oriented in any direction relative to the Earth). Thus, should a residual spurious effect remain after the elimination of glitches, it would not be confused with the signal. 

This analysis  applied to  \textit{Microscope} real measured data after removal of the outliers would establish beyond question the result of the experiment.

\section{Conclusions}\label{Sec:Conclusions}  

A space test of the weak equivalence principle  in the field of the Earth has enormous potential for a leap forward in precision, by building on a strong driving signal and better instrument isolation from nearby masses in space than in any ground laboratory, hence easier control of systematic effects from local disturbances (\cite{PerBraginskyPLA2018}). 

A satellite test of the WEP has been over 40 years in the making, a number of designs were proposed, one --\textit{Microscope}-- has made it to flight, and has  reported about two orders of magnitude improvement over the best laboratory experiments to date.  

The experiment appears to be limited by thermal noise from internal damping in the grounding wires whose stiffness was predicted incorrectly (by two orders of magnitude) in contrast with previous results from GOCE.  Once in orbit, in order to reduce the higher level of noise, the spin rate of the satellite was raised to a considerably larger rate than planned, and  an unexplained  noise jump was oserved.  We find jumps, up or down, also in sequential sessions at the same spin rate, which are inexplicable on the basis of the current knowledge of internal damping noise and the physical quantities it depends upon.

We find that  voltage fluctuations at sub-$\mu\rm V$ level, coupled to the large bias voltage made necessary by the capacitive control scheme, may cause the detected acceleration to leap in the way revealed by the data, and we suggest a potential cause in erratic patch charge rearrangements leading to an unstable zero. An issue to be taken into account regardless of the gold wire.

The experiment was affected by a large number of glitches, anomalous releases of energy, originating in the spacecraft, producing large acceleration differences at the  signal and spin frequencies, their harmonics and their difference. We argue the form of such glitches supports the hypothesis they were triggered by cycles of heating and cooling, by the Sun and Earth, of the loose insulation foils. We find evidence that the way the glitches  were treated --removed and replaced with artificial reconstructed data-- leaves extant effects at the critical frequencies, including the frequency of the signal, that could mimic a violation signal or cancel an effect (systematic error or real signal). We argue that an analysis as successfully carried out in the  \textit{E\"ot-Wash}  laboratory tests  would be tolerant of the missing data without affecting the sensitivity of the experiment.

Any future WEP experiment in space must  prove by its very design that it will not suffer from glitches (GOCE showed the way) and must avoid  introducing artificial data potentially compromising the analysis.

Our investigation, having identified the strong and weak points of the first space test of the WEP, shows the way to the desirable properties of an experiment design targeting yet another leap in precision, to $10^{-17}$ or better: differential instrument; well-defined and stable zero point; large gaps between test masses; rotation about a symmetry axis; 2D sensitivity in the plane of the signal; fast spin.

 \href{https://doi:10.1103/PhysRevLett.125.191101}{Physical Review Letters 125(191101):1--5.}


\begin{acknowledgements}
  Thanks are due to Giuseppe Catastini for many discussions on the heritage  from  ESA's GOCE mission. While at Alcatel Alenia Space (now Thales Alenia Space), GC was the chief scientist responsible for the design, development and operation of the GOCE end-to-end software simulator, a crucial tool for a mission the performance of which could not be fully tested on the ground.    
\end{acknowledgements}

\section*{Conflict of Interest}
The authors declare that they have no conflict of interest.

\end{document}